# Evolutionary simulations of autopoietic cells with cognition

Hirotaka Matsufuji and Osamu Narikiyo*

*Department of Physics, Kyushu University, Fukuoka 819-0395, Japan*

ABSTRACT

The minimal requirements for life are autopoiesis and cognition. We propose autopoietic models with cognition and perform three classes of evolutionary simulation. In our models the plasticity of the metabolic cycle and the regulation function of the membrane are the bases for the cognition. The cognitive cells show the adaptation and the evolution. The environment also shows the evolution via the interaction with the system of cells. This is a prototype of the co-evolution of the living system and its environment.

\* Corresponding author.
 *E-mail address:* narikiyo@phys.kyushu-u.ac.jp (O. Narikiyo)

# 1. Introduction

After Luisi (Luisi 2003, Bitbol and Luisi 2004, Luisi 2006) we assume that the minimal requirements for life are autopoiesis and cognition.

*A system is autopoietic if:*
*(a) it has a semi-permeable boundary,*
*(b) the boundary is produced from within the system, and*
*(c) it encompasses reactions that regenerate the components of the system.*

This definition written by Bourgine and Stewart (Bourgine and Stewart 2004) is quoted from Luisi's summary (Luisi 2003) of Varela's book (Varela 2000). We adopt this definition of autopoiesis.

Depending on the scope of the theories the definition of cognition differs. In this paper we adopt the definition written by Bourgine and Stewart (Bourgine and Stewart 2004):

*management of the interactions between an organism and its environment.*

We think that the most primitive implementation of the *management* consists of two elements. One is a function of the cell membrane regulating the flow of materials between inside and outside of the cell. The other is a plasticity of the metabolic cycle in the cell.

Most studies of autopoiesis for cellular life based on computer simulations focus on the establishment of a single cell (McMullin and Varela 1997, Suzuki and Ikegami 2007). However, such studies lack the viewpoint of the evolution. We think that the evolvability is one of the important elements of life and that a cognitive system acquires the evolvability. To discuss the evolution we formulate a computer simulation of the group of autopoietic cells with cognition.

We have performed three classes of evolutionary simulation: the model employed in the section 3 stresses the aspect of adaptation, the section 4 the aspect of evolution, and the section 5 the aspect of co-evolution.

# 2. Cells

If we study the autopoietic behavior of a single cell, we have to show the establishment of the metabolic cycle and the membrane (McMullin and Varela 1997, Suzuki and Ikegami 2007). On the other hand, we are interested in the evolution of *already* autopoietic cells. Thus we start our simulation with *already* autopoietic cells. Since our concern is the behavior of the group of cells, the description of each cell is chosen to be simple as discussed in the following sections.

To discuss the cognitive behavior we add the following two assumptions to ordinary autopoietic cells.

*(1) plasticity of metabolic cycle*

The core of a cell in our simulation is a metabolic cycle under the flow of energy and chemical materials through its membrane. Although we do not implement it explicitly, we assume a genetic or epigenetic regulation of the cycle. The regulation here is a very primitive one expected for very early cells (Nishio and Narikiyo 2013). But we think it is necessary for the stability of the cycle. Via the interaction of a cell to the other cells or to the environment, turnovers among stable cycles can occur. This is our assumption of the plasticity of the metabolic cycle.

*(2) regulation function of membrane*

The membrane is one of the important ingredients of an autopoietic cell. It is assumed to be a semi-permeable boundary in previous studies. Here we add a function to the membrane to regulate the flow of materials between inside and outside of the cell. Although the implementation of such a function needs some protein-machinery embedded in the membrane (Mann 2008), we skip the implementation and assume the regulation function of the membrane.

## 3. Adaptation

Our simulation resembles an experiment of the evolutionary molecular engineering where various kinds of molecules under the flow of energy and materials are selected by their fitness. Although the actual experiment is described by reaction-diffusion equations, we neglect the spatiotemporal effect of the diffusions and only consider the reactions in the following.

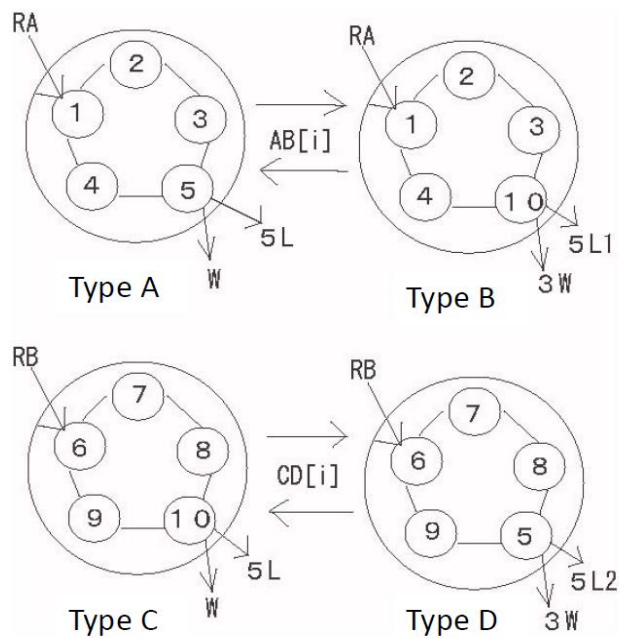

**Fig. 1** Metabolic cycles, A, B, C, and D. These cycles are put in a cell whose membrane consists of the elements, L, L1, and L2.

First we prepare 100 cells. Each cell has 100 substrates which might constitute metabolic cycles. We consider 10 kinds of substrates: X1, X2, X3, X4, X5, X6, X7, X8, X9, and X10. These 10 kinds of substrates are distributed randomly in each cell.

As shown in Fig. 1 we consider only 4 kinds of cycles which link 5 kinds of substrates. Every reaction links these substrates are assumed to be stabilized and strengthened by enzymes under a genetic or epigenetic regulation. Type-A cycle links X1, X2, X3, X4, and X5. Type-B cycle links X1, X2, X3, X4, and X10. Type-C cycle links X6, X7, X8, X9, and X10. Type-D cycle links X6, X7, X8, X9, and X5. Type-B is the mutant of Type-A. Type-D is the mutant of Type-C. We neglect the other reactions than these 4 cycles.

The strength of a cycle is expressed by the minimum number of the substrates which constitute the cycle.

In the initial state the allowed types of the cycle are A and C. Namely there are no mutants in the initial state. The allowed types are specified by the order parameters in our simulation. The order parameter is a simplified measure of the status of the regulation network of the cycle. The network is only assumed but not implemented in our simulation. We assume that the regulation is genetically or epigenetically maintained and that the status is inherited through the cell division.

The order parameter $AB[i]$ for the $i$-th cell represents the status between the cycles A and B. If $0 \leq AB[i] \leq 0.5$, the cycle A is allowed and B is inhibited. If $0.5 < AB[i] \leq 1$, the cycle B is allowed and A is inhibited. Namely the cycles A and B are mutually exclusive. In the initial state the distribution of $AB[i]$ is chosen to be the Gaussian whose center is 0.25 and variance is 0.02. We use the Gaussian shape only for $0 \leq AB[i] \leq 0.5$ and put all the initial value $AB[i]$ within this range.

The order parameter $CD[i]$ for the $i$-th cell represents the status between the cycles C and D. If $0 \leq CD[i] \leq 0.5$, the cycle C is allowed and D is inhibited. If $0.5 < CD[i] \leq 1$, the cycle D is allowed and C is inhibited. Namely the cycles C and D are mutually exclusive. In the initial state the distribution of $CD[i]$ is chosen to be the Gaussian whose center is 0.125 and variance is 0.005. We use the Gaussian shape only for $0 \leq CD[i] \leq 0.25$ and put all the initial value $CD[i]$ within this range.

The time-steps of the simulation increase by one when resource materials are supplied simultaneously to every cell. The input to the cycles A and B is the resource material RA. The input to the cycles C and D is the resource material RB. When one resource material is supplied to an allowed cycle, every number of the substrates belonging to the cycle increases by one. The output is the element of the membrane (L, L1, L2) and the waste material (W). We only consider the flow of materials but not the flow of energy. The cycles A and C produce the membrane element L. The cycle B produces L1. The cycle D produces L2. Every cycle produce 5 membrane elements per one resource material. The cycles A and C push out one waste material per one resource material. The cycles B and D push out 3 waste materials per one resource material.

We assign the following functions to the membrane elements. The element L only regulates the input: the number of resource material RA coming into the cell through the membrane is Int[#L / 100] for one time-step where #L is the number of L at the end of the preceding time-step and Int[ $x$ ] represents the integer part of the real number $x$. The element L1 regulates the input: the number of resource material RA coming into the cell is Int[#L1 / 20]. Namely L1 is more effective than L by five times to get resource. For example, Int[#L1 / 20] = 2 when #L1 = 50. The resource material RB is treated by the same manner. The element L2 discharges waste materials: the number of waste material W going out of the cell is 3 times Int[#L2 / 20] for one time-step.

Since the products L1 and L2 are supposed to be more complex than L, the amount of waste material coming from the cycles B and D are set to be 3 times of that from A and C.

At the end of the time-step, if the summation of the numbers of membrane elements, #L + #L1 + #L2, in a cell becomes 200, we divide such a cell into two cells. The compositions of the membrane elements of the daughter cells are set to be the same as that of the mother cell as possible. The numbers of waste materials of the daughter cells are also set to be the same as possible. Each substrate of the metabolic cycle is distributed to either daughter cell with equal probabilities.

At the cell division the values of $AB[i]$ and $CD[i]$ can change by $+0.1$ or $-0.1$. This process represents the random mutation of the genetic or epigenetic regulation. The sign of the change is randomly chosen with equal probabilities. The change is accepted if the changed value is in the range between 0 and 1 but rejected if the changed value is out of this range. In the latter case the values of $AB[i]$ and $CD[i]$ stay through the cell division.

Just after the cell division we delete unhealthy cells which have too many waste materials: the criterion for the deletion is #W / #X > 0.4 where #W is the number of waste material and #X is the summation of the numbers of all the substrates in the daughter cell. If the total number of cells exceeds 100 after this deletion, we delete excess cells randomly form all mother and daughter cells until the number becomes 100. Then the simulation after the cell division and the deletion starts with 100 cells. This restriction mimics the evolutionary molecular engineering experiment with a finite-volume pool of molecules. For large time-steps the unhealthy cells without the cycle D are plagued by waste materials and deleted.

The numbers #X and #W increase as the time-steps increase but the number of membrane elements for a cell is bounded as #L + #L1 + #L2 ≤ 200. Thus such a condition is unrealistic for large time-steps. But we have neglected this bad condition for simplicity.

Our conditions for the simulation favor the cycles B and D over A and C. The increase of the strength of the cycle B enhances the rapidity of the cell division. The increase of the cycle D enhances the removal of waste material. The probability of the survival against the selection increases by these enhancements. Namely the adaptability increases by these enhancements. Thus our simulation is biased to show an adaptation to a given condition: the adaptability of the cells increases as the time-steps increase.

As shown in Fig. 2 the number of cells with B and D cycles increases as the time-steps increase. This is an adaptation.

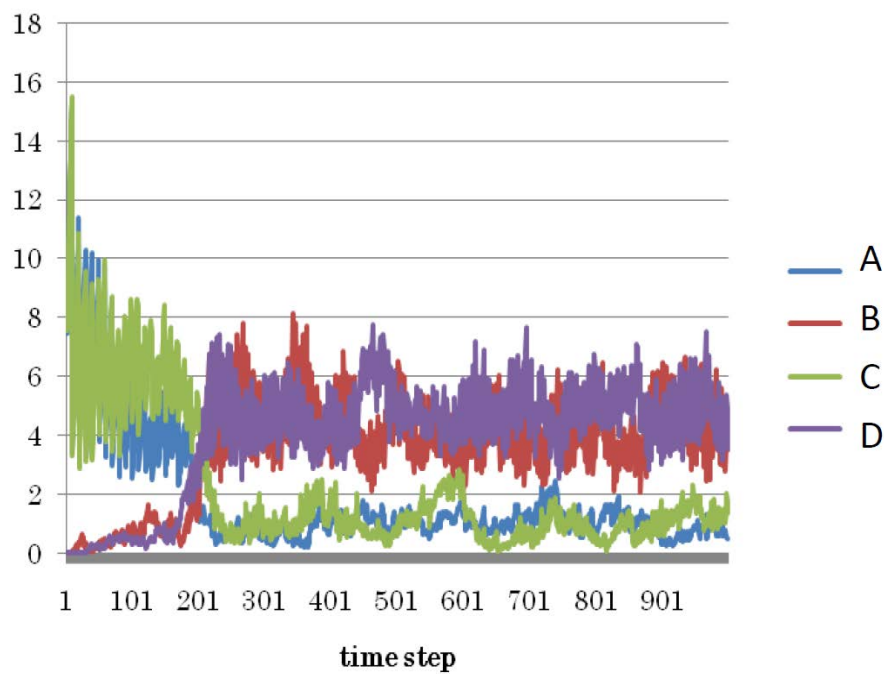

**Fig. 2**  The strength of the metabolic cycle per cell averaged over survived cells. The strength is defined by the minimum number of the substrates which constitute the allowed cycle.

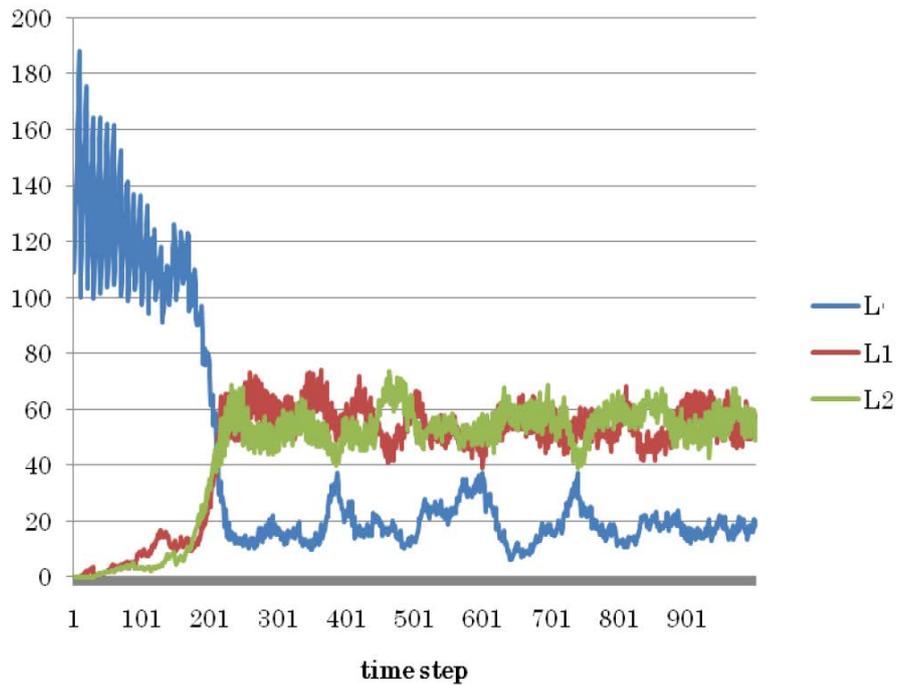

**Fig. 3**　The number of membrane element per cell averaged over survived cells.

As shown in Fig. 3 the ratios of membrane elements L1 and L2 increases as the time-steps increase. This is an adaptation.

## 4. Evolution

Next we perform a simulation with larger number of cells. Thus we simplify the description of a single cell.

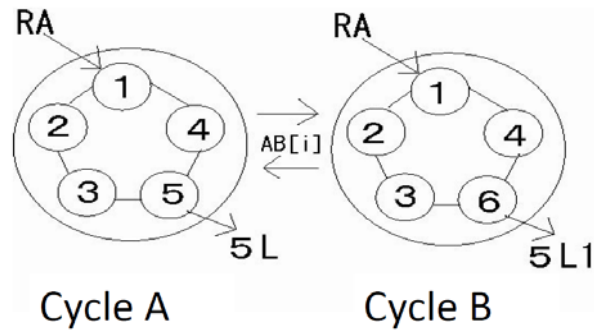

**Fig. 4**　Metabolic cycles, A and B. These cycles are put in a cell whose membrane consists of the elements, L and L1.

In the initial state we prepare 100 cells. Each cell has 100 substrates which might constitute metabolic cycles. We consider 6 kinds of substrates: X1, X2, X3, X4, X5, and X6. These 6 kinds of substrates are distributed randomly in each cell.

We consider the turnover between the cycles A and B. The definitions of the cycles A and B are almost the same as those in the previous section but we neglect the waste material here as shown in Fig. 4.

The initial condition and the time evolution for the order parameter $AB[i]$ are the same as those given in the previous section.

The efficiency of the membrane elements L and L1 is set to be as follows. The number of resource material RA coming into the cell through the membrane element L is Int[#L / 100] for one time-step. The number through L1 is Int[#L1 / 10]. Namely L1 is more effective than L by 10 times to get resource.

The rule for the cell division from mother cell to daughter cells is the same as that in the previous section. While we restrict the number of cells not to exceed 100 in the previous section, here we continue the simulation until the total number of cells exceeds 30000.

Here we define the MC (Monte Carlo) step in our simulation. (0): First we count the total number of cells #Cell. Next we supply resource materials to one cell randomly chosen from all mother and daughter cells. The number of RA supply to the chosen cell is determined by Int[#L / 100] + Int[#L1 / 10]. If this supply procedure is repeated #Cell times, one MC step ends. Then we increase the number of the MC step by one, go back to (0), and repeat these procedures.

Every 10 MC steps we delete inferior cells which are poor in getting resource materials. We name the criterion for the deletion the selection pressure $SP$. Such a selection pressure is naturally expected in a situation where the number of available resource materials is limited. At the time of the selection we check the number of resource material $RA[i]$ supplied during the past 10 MC steps for each cell. If $RA[i] < SP$, the $i$-th cell is deleted.

Although it is an expected result by the organization of the simulation, the larger the selection pressure is, the faster the evolution progresses as shown in Fig. 5. Here we interpret the evolution as getting larger adaptability. The adaptability of a cell is estimated by the efficiency of the membrane, $\#L + 10 \times \#L1$.

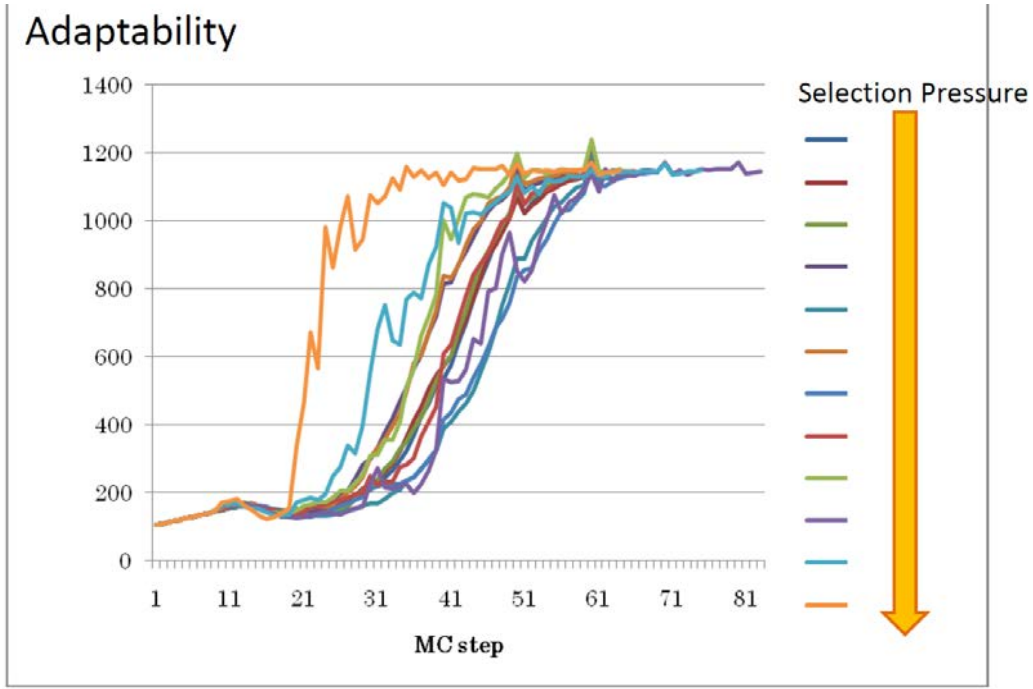

**Fig. 5**  Adaptability, #L + 10×#L1, averaged over survived cells against the selection. The selection pressure is varied as $SP = 1,2,3,4,5,6,7,8,9,10,11,12$.

## 5. Co-evolution

  In this section we also consider the change of the environment of the cells. In the initial state we prepare #RA of resource materials. The summation of the numbers of the resource and the waste in the environment is kept to be #RA throughout the simulation.

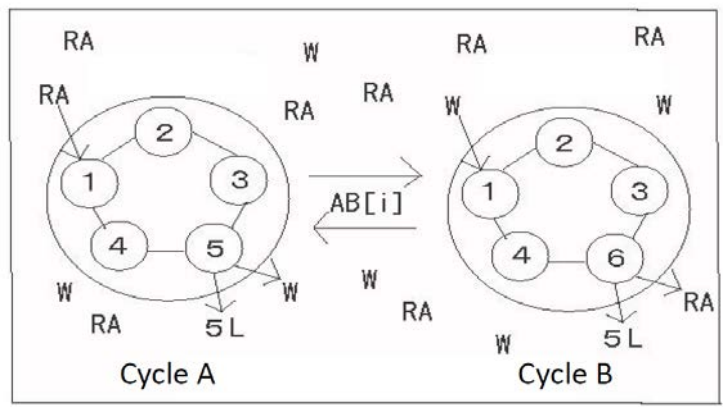

**Fig. 6**  Metabolic cycles, A and B. The roles of resource and waste for A are reversed for B. These cycles are put in a cell whose membrane consists of the element, L.

In the initial state we prepare 100 cells. Each cell has 100 substrates which might constitute metabolic cycles. We consider 6 kinds of substrates: X1, X2, X3, X4, X5, and X6. These 6 kinds of substrates are distributed randomly in each cell.

We consider the turnover between the cycles A and B. The definitions of the cycles A and B are almost the same as those in the section 3 but the resource and the waste are reversed for the cycle B as shown in Fig. 6.

The membrane of the cell with the cycle A absorbs one RA and discharges one W per one time-step. The membrane of the cell with the cycle B absorbs one W and discharges one RA per one time-step. The cells with A and the cells with B are in co-existence and co-prosperity. The summation of the numbers of the resource and the waste in the environment is unchanged throughout the simulation.

The initial condition and the time evolution for the order parameter $AB[i]$ are the same as those given in the section 3.

The rule for the cell division from mother cell to daughter cells is the same as that in the section 3. Just after the cell division we delete excess cells randomly form all mother and daughter cells until the total number of cells becomes 100. This random deletion is the same as that employed in the section 3. Then the simulation after the cell division and the deletion starts with 100 cells.

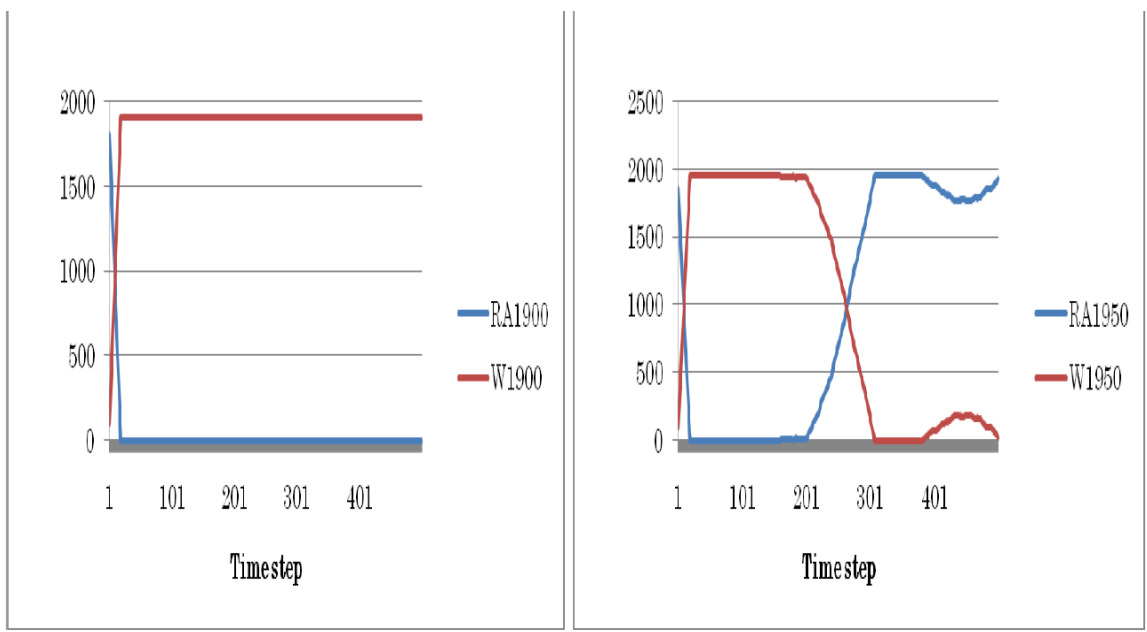

**Fig. 7**  The numbers of RA and W in the environment. The initial value is set as #RA=1900 in the left and #RA=1950 in the right.

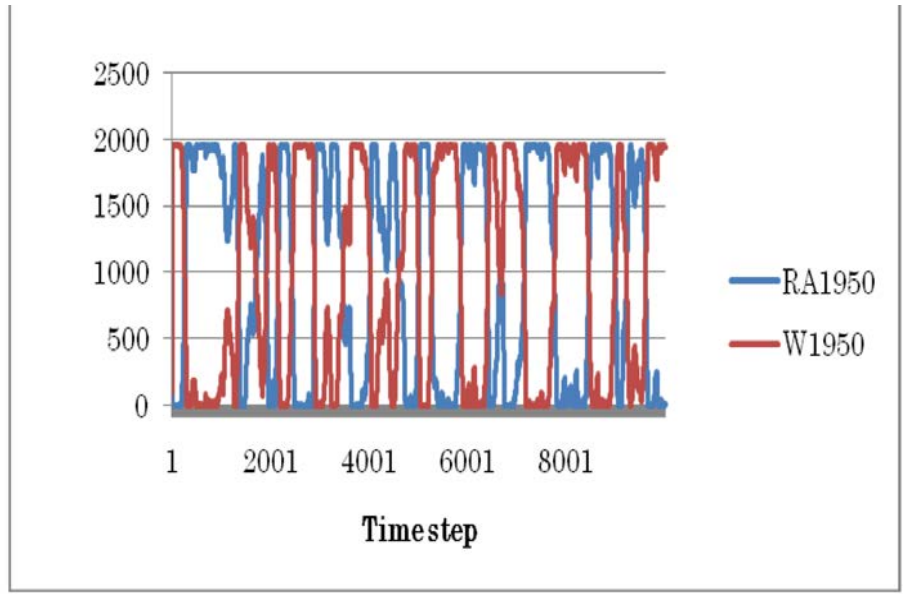

**Fig. 8**　The numbers of RA and W in the environment. The initial value is set as #RA=1950.

As shown in Fig. 7 the co-existence of the type-A and type-B cells is seen in the case of #RA=1950 but not in the case of #RA=1900. Since RA is exhausted before the genetic drift leads to type-B cells from type-A cells, type-B cells do not appear for #RA<1950.

The numbers of type-A cells and type-B cells oscillate as seen from Fig. 8. For the simulations for #RA>1950 a similar time-dependence to that in Fig. 8 is seen.

We have observed that corresponding to the evolution of the cell type from A to B the composition of the environment changes. This is a prototype of the co-evolution of the living system and its environment.

## 6. Conclusion

We have proposed autopoietic models with cognition and performed three classes of evolutionary simulation. In our models the plasticity of the metabolic cycle and the regulation function of the membrane are the bases for the cognition.

The cognitive cells have shown the adaptation and the evolution.

The environment also has shown the evolution via the interaction with the system of cells. This is a prototype of the co-evolution of the living system and its environment.